\documentclass[twocolumn]{aastex62}

\shorttitle{X-ray variability changes and jet ejection in MAXI J1820+070}
\shortauthors{Homan et al.}

\newcommand{\maxi}{MAXI J1820+070}

\begin{document}

\title{A rapid change in X-ray variability and a jet ejection in the black hole transient MAXI J1820+070}

\correspondingauthor{Jeroen Homan}
\email{jeroenhoman@icloud.com}

\author{Jeroen Homan}
\affiliation{Eureka Scientific, Inc., 2452 Delmer Street, Oakland, CA 94602, USA}
\affiliation{SRON, Netherlands Institute for Space Research, Sorbonnelaan 2, 3584 CA Utrecht, The Netherlands}

\author{Joe Bright}
\affiliation{Astrophysics, Department of Physics, University of Oxford, Denys Wilkinson Building, Keble Road, Oxford OX1 3RH, UK}

\author{Sara E. Motta}
\affiliation{Astrophysics, Department of Physics, University of Oxford, Denys Wilkinson Building, Keble Road, Oxford OX1 3RH, UK}

\author{Diego Altamirano}
\affiliation{School of Physics and Astronomy, University of Southampton, Southampton, SO17 1BJ, UK}

\author{Zaven Arzoumanian}
\affiliation{X-ray Astrophysics Laboratory, Astrophysics Science Division, NASA Goddard Space Flight Center, Greenbelt, MD 20771, USA}

\author{Arkadip Basak}
\affiliation{Anton Pannekoek Institute, University of Amsterdam, Science Park 904, 1098 XH Amsterdam, The Netherlands}

\author{Tomaso M.\ Belloni}
\affiliation{INAF - Osservatorio Astronomico di Brera, via E. Bianchi 46, I-23807 Merate, Italy}

\author{Edward M.\ Cackett}
\affiliation{Department of Physics and Astronomy, Wayne State University, 666 W. Hancock Street, Detroit, MI 48201, USA}

\author{Rob Fender}
\affiliation{Astrophysics, Department of Physics, University of Oxford, Denys Wilkinson Building, Keble Road, Oxford OX1 3RH, UK}

\author{Keith C.\ Gendreau}
\affiliation{X-ray Astrophysics Laboratory, Astrophysics Science Division, NASA Goddard Space Flight Center, Greenbelt, MD 20771, USA}

\author{Erin Kara}
\affiliation{MIT Kavli Institute for Astrophysics and Space Research, MIT, 70 Vassar Street, Cambridge, MA 02139, USA}

\author{Dheeraj R.\ Pasham}
\affiliation{MIT Kavli Institute for Astrophysics and Space Research, MIT, 70 Vassar Street, Cambridge, MA 02139, USA}

\author{Ronald A.\ Remillard}
\affiliation{MIT Kavli Institute for Astrophysics and Space Research, MIT, 70 Vassar Street, Cambridge, MA 02139, USA}

\author{James F.\ Steiner}
\affiliation{Harvard-Smithsonian Center for Astrophysics, 60 Garden Street, Cambridge, MA 02138, USA}

\author{Abigail L.\ Stevens}
\affiliation{Department of Physics \& Astronomy, Michigan State University, 567 Wilson Road, East Lansing, MI 48824, USA}
\affiliation{Department of Astronomy, University of Michigan,1085 South University Avenue, Ann Arbor, MI 48109, USA}

\author{Phil Uttley}
\affiliation{Anton Pannekoek Institute, University of Amsterdam, Science Park 904, 1098 XH Amsterdam, The Netherlands}

\begin{abstract}

We present {\it Neutron Star Interior Composition Explorer} X-ray and Arcminute Microkelvin Imager Large Array radio observations of a rapid hard-to-soft state transition in the black hole X-ray transient \maxi. During the transition from the hard state to the soft state a switch between two particular types of quasiperiodic oscillations (QPOs) was seen in the X-ray power density spectra, from type-C to type-B, along with a drop in the strength of the broadband X-ray variability and a brief flare in the 7--12 keV band. Soon after this switch ($\sim$1.5--2.5 hr) a strong radio flare was observed that corresponded to the launch of superluminal ejecta. Although hints of a connection between QPO transitions and radio flares have been seen in other black hole X-ray transients, our observations constitute the strongest observational evidence to date for a link between the appearance of type-B QPOs and the launch of discrete jet ejections.

\end{abstract}


\section{Introduction}

Black hole low-mass X-ray binaries (BH LMXBs), in which a stellar-mass black hole accretes matter from a low-mass companion star, are predominantly transient systems that spend most of their time in a low-luminosity (highly sub-Eddington) quiescent state. During their occasional outbursts the X-ray luminosity of BH LMXBs can increase by factors of up to $10^6$, often showing transitions between various X-ray spectral/variability states in the process. Several classification schemes have been proposed for these states \citep[e.g.,][]{hobe2005,remc2006,bemo2016}. Two main stable spectral states are usually recognized: the hard state, which shows strong rapid (subsecond) X-ray variability and a spectrum dominated by Comptonized power-law emission, and the soft state, in which rapid X-ray variability is weak and where a thermal disk blackbody component dominates the spectrum. The transitions between these two spectral states, during which sources are often classified as being in an intermediate state, are accompanied by strong evolution of the X-ray variability properties. As a source starts the transition from the hard state to the soft state, strong (type-C) low-frequency quasi-periodic oscillations (QPOs) are usually present in the power density spectra \citep[see][for the various low-frequency QPO types]{wihova1999,resomu2002,cabest2005}. Their frequencies increase from $\sim$0.01 Hz to $\sim$10 Hz as the spectrum softens. At some point during the transition the type-C QPOs and the associated strong band-limited noise are (temporarily) replaced by other types of QPOs (type-A or type-B, both in $\sim$4--9 Hz range) and weaker band-limited noise. While the type-C QPOs are commonly interpreted as being the result of relativistic precession of the inner accretion flow \citep[e.g.,][]{schomi2006,indofr2009}, the nature of the other two types is less clear. Based on their dependence on viewing angle \citep[stronger in more face-on sources,][]{mocahe2015} and their proximity in time to jet launches \citep[][see next paragraph]{sobeca2008}, the type-B QPOs have been suggested to be related to the production of jet(-like) outflows, but no physical model has been proposed.

Within a few days of the (occasionally rapid) changes between the various types of low-frequency QPOs, radio flares  have been observed in some BH LMXBs \citep[see, e.g.,][]{cokaja2001,brfemc2002,gacofe2004,misial2012,rutemi2019}. The flares last hours to days and high angular-resolution radio and X-ray observations have shown that  they are the result of discrete ejection events  that occurred prior to the radio flares \citep[][and references above]{cofetz2003,cokafe2005}.  Suggestions were made that the ejection events and the resulting radio flares are associated with the appearance of type-B QPOs \citep[e.g.][]{sobeca2008}, but detailed studies of X-ray variability and radio flares in XTE J1550$-$564, GX 339$-$4, XTE J1859+226, H1743$-$322, and MAXI J1535--571 were inconclusive as to whether such a relation exists \citep{fehobe2009,misial2012,rutemi2019}. These studies were all hampered by gaps in either the X-ray and/or radio coverage, with the durations of these gaps ranging from $\sim$10 hr to $\sim$4 days, preventing an unambiguous association of the radio flares and ejection events with the appearance/disappearance of any of the low-frequency QPO types to better than half a day to a few days, depending on the source. In particular, the moment at which the strong change in rapid X-ray variability occurred was not observed in any of the above sources; it must have taken place during a gap in the X-ray coverage. Also, only in a two cases, GX 339-4 and XTE J1859+226, was the onset of the radio flare  observed, but the flares were preceded by gaps in the X-ray coverage of $\sim$2 days and $\sim$16 hr, respectively. In the other cases only part of the rise was covered or only (part of) the decay phase was observed. 

In this Letter, we present observations of a radio flare that occurred during a hard-to-soft state transition in MAXI J1820+070. MAXI J1820+070 is an X-ray transient that was discovered with the Monitor of All-Sky X-ray Image (MAXI) in 2018 March  \citep{kaneyo2018}. It was soon realized \citep{de2018} that MAXI J1820+070 was the X-ray counterpart of the optical transient ASASSN-18ey, discovered a few days earlier \citep{tushho2018b}. Observations at various wavelengths soon suggested that the source is a BH LMXB \citep{barule2018,brfemo2018,dese2018,utgema2018}. The black hole nature of the source was confirmed with a dynamical mass measurement by \citet{tocaji2019}, who put the mass in the 7--8 $M_\odot$ range. \maxi\ is a relatively nearby BH LMXB, with a distance of 2.96$\pm$0.33 kpc \citep{atmiba2020},  and it has low interstellar absorption  \citep[$\sim2\times10^{21}$  atoms\,cm$^{-2}$;][]{kamosa2019}. After spending roughly the first half of its 8 month outburst in the hard state,  MAXI J1820+070 underwent a rapid state transition to the soft state in early July of 2018 \citep{houtge2018a,houtge2018b}. During this transition a fast switch from type-C to type-B QPOs was observed. Contemporaneous radio observations revealed a strong radio flare that was the result of the launch of superluminal ejecta \citep{brmofe2018,brfemo2020}  and that, as we show in this Letter, may have started within  $\sim$2--2.5 hr of the QPO transition. This presents the strongest observational link between the two phenomena to date.

\section{Observations and data analysis}\label{sec:obs}

\subsection{{\it NICER}}

The {\it Neutron Star Interior Composition Explorer} \citep[{\it NICER};][]{gearad2016} is an X-ray timing instrument on board the {\it International Space Station}. {\it NICER} consists of 56 concentrators, each coupled to a focal plane module (FPM) that houses a silicon drift detector. Of these, 52 were functional at the start of {\it NICER} operations in 2017 July.  {\it NICER} provides coverage in the soft (0.2--12 keV) X-ray band with CCD-like spectral resolution and time-tags photons with an absolute timing precision of $\sim$100 ns. {\it NICER} observed the 2018 outburst of MAXI J1820+070 in great detail. Between 2018 March 5 and 2018 November 21 the source was observed with a total of 750 individual visits and an exposure time of $\sim$566 ks (after data filtering and cleaning). All observations were reprocessed with {\tt nicerl2} (processing version 20180226), which is part of the {\it NICER} software tools in HEASOFT v6.25, using standard filtering criteria \citep[see, e.g.,][]{stutal2018}. For each observation we extracted light curves in various energy bands with a time resolution of 16s. During the brightest phases of the outburst the number of active FPMs had to be lowered (in some cases to just 17) to avoid data dropouts caused by internal telemetry saturation. We  therefore normalized the light curves by the number of active FPMs. Light curves in the 13--15 keV band (where no source contribution is expected, due to {\it NICER}'s effective area) were used to remove segments of elevated background events (13--15 keV count rates $>$1.0 counts\,\,s$^{-1}$ per 52 FPM). The light curves were not background-subtracted, which is not a concern for the (high count-rate) phase of the outburst that we focus on. We also ran Fast Fourier Transforms on 16s time segments with a time resolution of 2048$^{-1}$ s to create power density spectra with a frequency range of 16$^{-1}$--1024 Hz. This was done for a variety of energy bands. 

\subsection{AMI-LA}

\maxi\ was observed in the radio with the Arcminute Microkelvin Imager Large Array \citep[AMI-LA][]{zwbabi2008,hirape2018} at a central frequency of 15.5 GHz, across a 5 GHz bandwidth consisting of 4096 channels. A total of 183 observations were made between  2018 March and 2018 December, with observation lengths between 1 hr and 8 hr. Data were reduced using the custom software pipeline {\tt REDUCE\char`_DC} \citep[e.g.,][]{dafrda2009,pescgr2015}, which flagged radio frequency interference and performed absolute flux and phase reference calibration. 3C 286 and J1824+1044 were used as the absolute flux and phase reference calibrator, respectively. We created a 30 minute time resolution light curve from observations taken between the 5\textsuperscript{th} and 9\textsuperscript{th} of July directly from the (channel and baseline averaged) \textit{uv}-amplitude coefficients. Errors were derived from the standard deviation of the amplitude coefficients within each time bin, combined in quadrature with a 5\% absolute flux scale uncertainty. Data points for which the error on flux density was larger than 25\% of the flux density itself were filtered out.

\section{Results}\label{sec:results}

\subsection{X-Ray evolution}\label{sec:xray}

\begin{figure}[t]
\includegraphics[width=8.5cm]{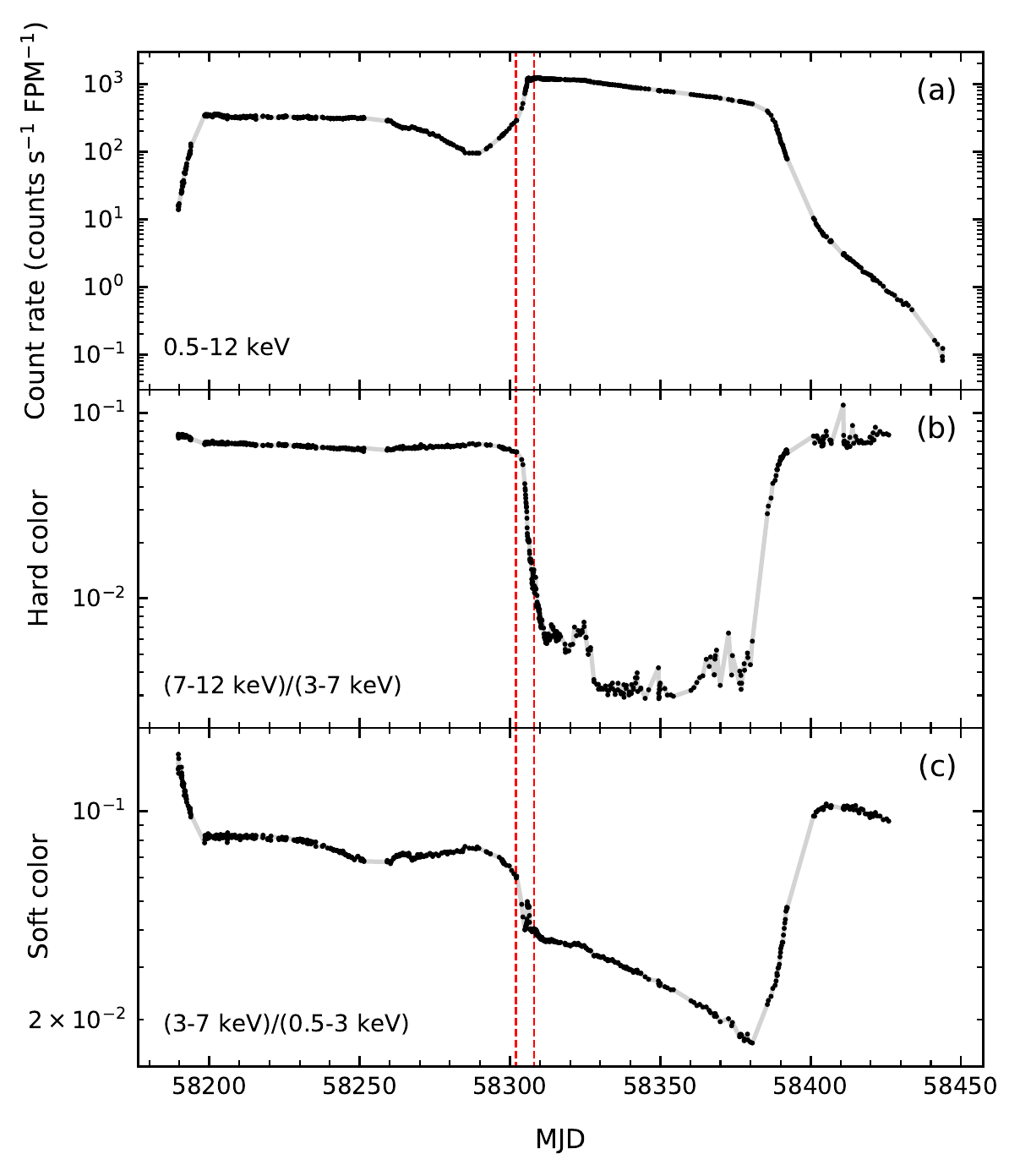}
\caption{Evolution of the {\it NICER} X-ray count rate and colors of \maxi\ during its 2018 outburst: (a) 0.5--12 keV light curve, (b) hard color, and (c) soft color. The definitions of the colors can be found within panels (b) and (c) (see also main text). Each data point represents the average of a single-orbit dwell. The red dashed lines mark part of the hard-to-soft state transition that is shown in detail in Figure \ref{fig:transition}. }\label{fig:outburst}
\end{figure}

\begin{figure*}[t]
\centerline{\includegraphics{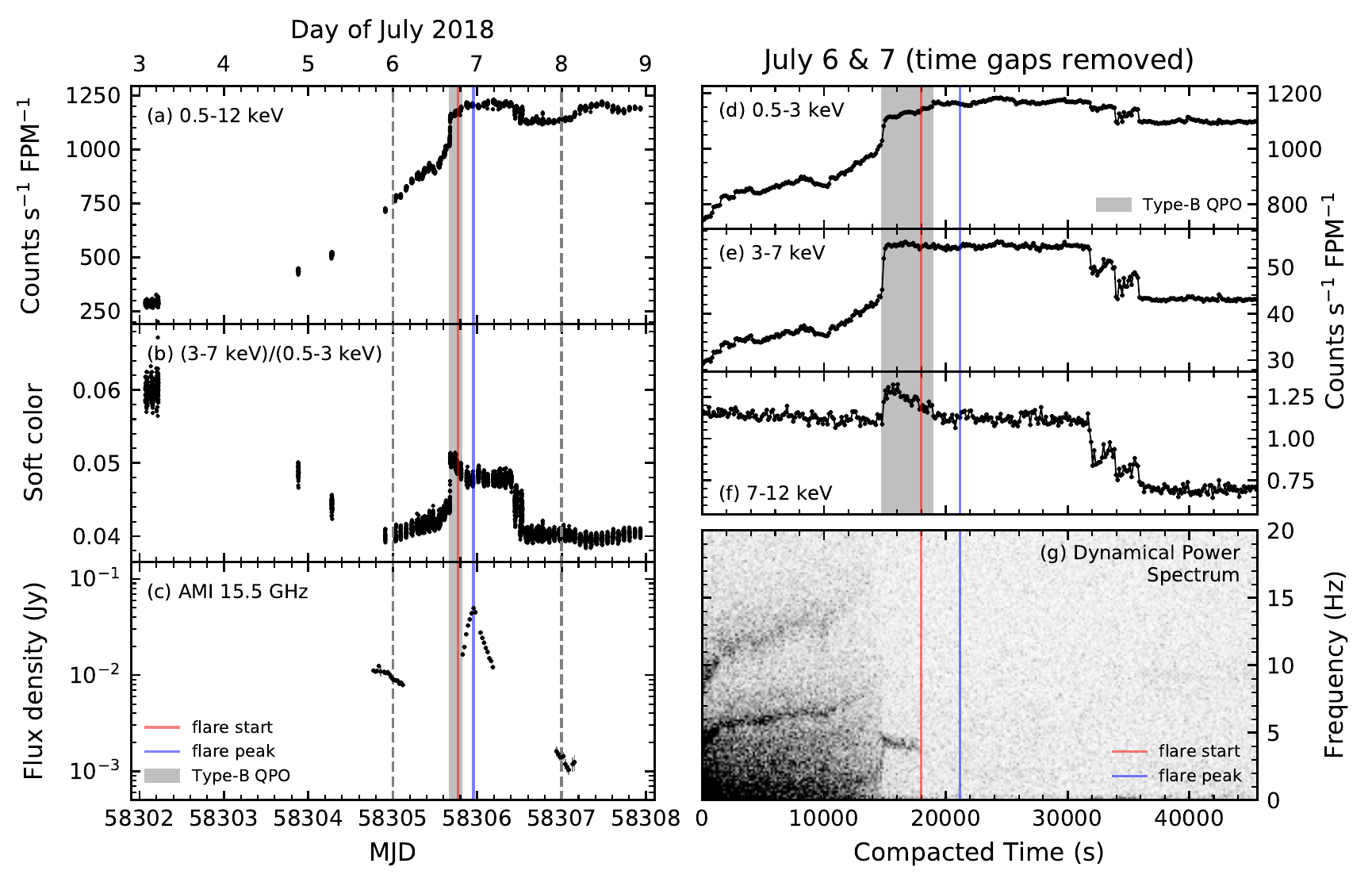}}
\caption{Detailed look at the rapid state transition in \maxi. Panels (a) and (b) show the 0.5--12 keV count rate and soft color with a time resolution of 16\,s. Panel (c) shows the AMI-LA 15.5 GHz radio light curve, with a time resolution of 30 minutes. The gray shaded areas in panels (a)--(f) mark the presence of the type-B QPO. In panels (d)--(g) only data taken on July 6 and 7 are shown (this period is marked with dashed vertical lines in panels (a)--(c). All time gaps were removed in panels (d)--(g) and data were rebinned to a time resolution of 128s - we refer to time in these panels as ``compacted time'', $t_c$. Panels (d)--(f) show light curves in the 0.5--3 keV, 3--7 keV, and 7--12 keV energy bands, respectively. Finally, in panel (g) we show a 0.3--12 keV dynamical power spectrum, with a frequency resolution of 0.125 Hz; the powers were not rms-normalized and Poisson-noise level was not subtracted. Darker shades correspond to stronger variability. The red vertical lines mark the start time of the radio flare (see Fig.\ \ref{fig:flare}), while the blue lines mark the peak of the radio flare.}\label{fig:transition}
\end{figure*}

In Figure \ref{fig:outburst}a we show the {\it NICER} light curve of the 2018 outburst of MAXI J1820+070 in the 0.5--12 keV band. Each data point represents the average for a single visit. In panels  \ref{fig:outburst}b and  \ref{fig:outburst}c we show two different X-ray colors; a hard color (ratio of count rates in the 7--12 keV and 3--7 keV bands) and a soft color (ratio of count rates in the 3--7 keV and 0.5--3 keV bands). While our definition of the hard color is similar to that of the soft color as frequently used with {\it Rossi X-ray Timing Explorer} ({\it RXTE}) data \citep[see, e.g.,][]{hoklro2003}, our soft color includes an energy range (0.5--3 keV) that was not covered by {\it RXTE}, making it much more sensitive to the thermal disk component in the spectra and to changes in the absorption. Note that we do not show the color evolution beyond MJD 58,426, as an increased relative background contribution started affecting the colors. 

Based on X-ray brightness and X-ray colors, we divide the outburst of \maxi\ into three parts: two hard state segments (roughly MJD 58,189--58,290 and MJD 58,400--58,443) that are separated by a period that includes two major state transitions and an extended soft-state phase. The high soft-color values during the rising hard state (MJD $<$58,195) are partly the result of the presence of a variable warm absorber that resulted in strong and spectrally hard dips \citep{hoalar2018,kamosa2019}. 

In panels \ref{fig:transition}a and \ref{fig:transition}b we show part of the hard-to-soft-state transition in more detail, with a time resolution of 16 s. This period is marked with red dashed lines in Figure \ref{fig:outburst} and corresponds to 2018 July 3--9 (MJD 58,302--58,308). As can be seen, in the days following July 3 the 0.5--12 keV count rate continued increasing (by a factor of more than 4), while the soft color continued to decrease.  By July 6 (MJD 58,305) the source began to deviate from this trend: the soft color started to increase, first slowly, then more rapidly, followed by an $\sim$1 day plateau. Around the same time the count rate entered a $\sim$1 day plateau as well. On July 7 both the count rate  and soft color showed sharp drops.

\begin{figure}[t]
\centerline{\includegraphics[width=8.5cm]{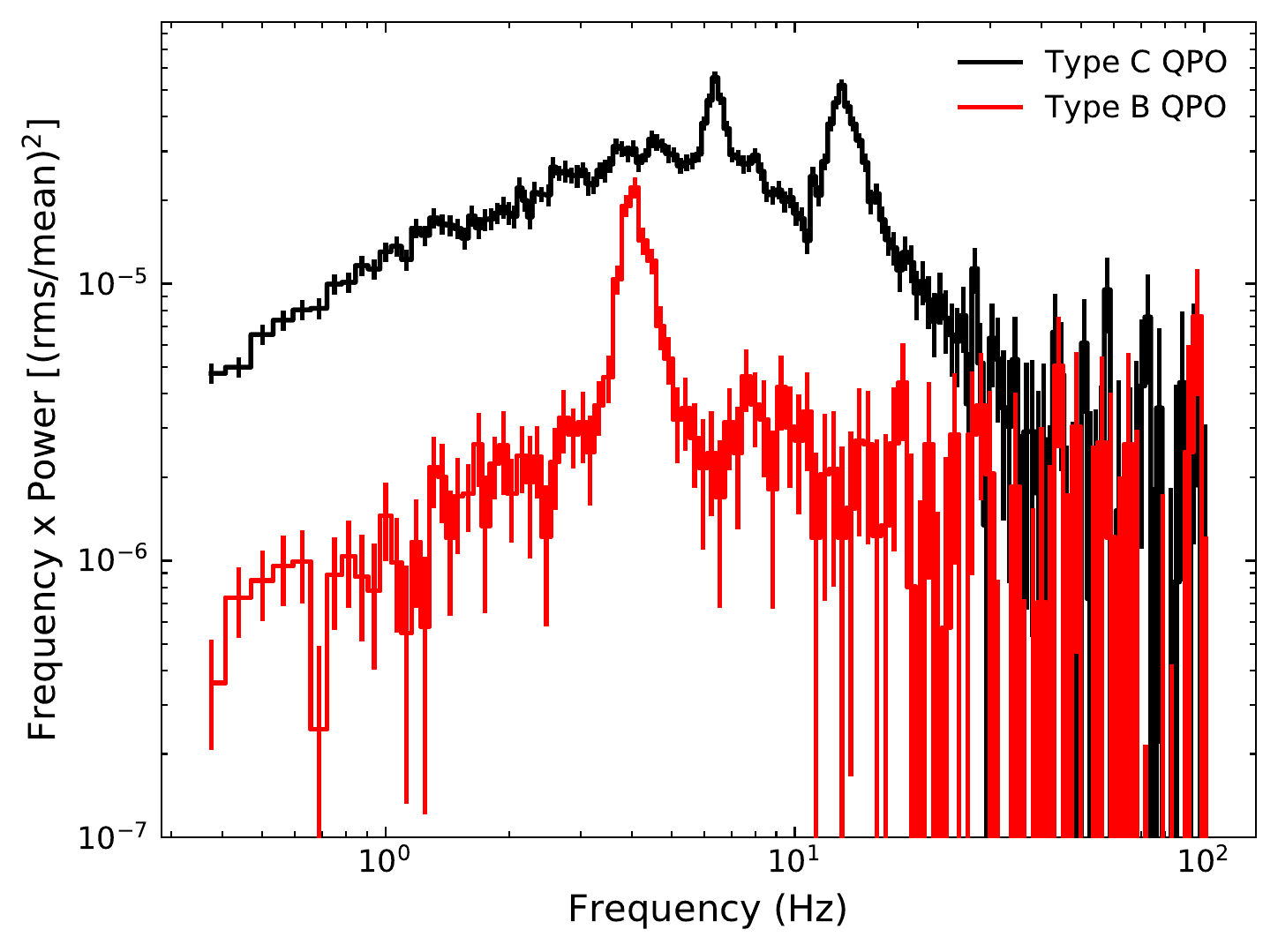}}
\caption{Two 0.3--12 keV power density spectra of \maxi. The black power spectrum is the average taken from the time interval $t
_c=2000-2700$\,s in panel \ref{fig:transition}g and shows type-C QPOs. The red power spectrum is the average taken from the time interval $t_c=16,000-17,600$\,s and shows a type-B QPO. Both time-averaged power spectra were rms-normalized \citep{beha1990} and the Poisson-noise level was subtracted.}\label{fig:qpos}
\end{figure}

Strong changes in the rapid X-ray variability were seen during this period of the outburst \citep{houtge2018a,houtge2018b}. In Figure \ref{fig:transition}g we show a 0.3--12 keV dynamical power spectrum from the observations made on July 6 and 7 (rebinned to a time resolution of 128 s and a frequency resolution of 0.125 Hz). Corresponding light curves in various energy bands are shown in panels  \ref{fig:transition}d--f. All time gaps were removed in panels \ref{fig:transition}d--g, to aid in the visualization of the evolution in the dynamical power spectrum. Since the x-axis in panels \ref{fig:transition}d--g no longer represents elapsed time, we will use the term `compacted time' ($t_c$) when referring to time marks in those panels.  Two harmonically related low-frequency QPOs, around 4.5 and 9 Hz, were present at the start of the dynamical power spectrum, along with strong noise at lower frequencies. A time-averaged power spectrum from the time interval $t_c = 2000$--$2700$\,s in panel \ref{fig:transition}g is shown in black in Figure \ref{fig:qpos}. Based on the shape and relative strength of the power spectral features, and taking into account the evolution prior to July 3 \citep{houtge2018a}, these QPOs can be identified as type-C.  Note that the overall strength of the variability in the 0.3--12 keV band (see Fig.\ \ref{fig:qpos})
 is considerably lower than typically seen in the {\it RXTE}/PCA band (2--60 keV) for similarly shaped power spectra. This is likely due to dilution by the accretion disk component at low energies.
The type-C QPOs increased in frequency (up to $\sim$8 and $\sim$16 Hz) and became weaker with time; they were no longer visible in the dynamical power spectrum around $t_c\approx13,000$\,s, but remained visible in time-averaged power spectra until $t_c\approx14,800$\,s.  The low frequency noise became weaker with time as well, by a factor of $\sim$2, and an additional broad peak around 1--2 Hz appeared to develop (not clearly visible in the dynamical power spectrum).

At $t_c\approx14,800$\,s several dramatic changes were observed. This time corresponds to MJD 58,305.67740 (July 6 16:15 UTC). In the light curves the count rates increased sharply. The jump in count rate was more pronounced at higher energies, as can be seen from panels  \ref{fig:transition}d--f, and resulted in the sharp increase in soft color that is visible in panel  \ref{fig:transition}b. At the same time the low-frequency noise became even weaker and the peaked noise feature around  1--2 Hz in the power spectrum appeared to evolve rapidly into a QPO around 4.5 Hz (see panel  \ref{fig:transition}g). A time-averaged power spectrum from the interval $t_c=16,000-17,600$\,s is shown in red in Figure \ref{fig:qpos}. Based on the QPO and noise properties, this QPO can be classified as a type-B QPO. The type-B QPO was only present for five good-time intervals ($t_c\approx4100$\,s in panel \ref{fig:transition}g) during which it slightly decreased in frequency, to $\sim$3 Hz. The duration of its presence is marked with a gray shaded area in panels \ref{fig:transition}(a-f). Toward the end of this time interval the QPO could no longer be seen in the dynamical power spectrum, due to a combination of intrinsic weakening of the QPO and a drop in the number of active detectors. Interestingly, a comparison with panel \ref{fig:transition}f suggests that the type-B was only present during the small flare in the 7--12 keV light curve that started around $t_c\approx14,800$\,s. After the type-B QPO disappeared ($t_c\approx18,900$\,s; MJD$\approx$58,305.81136, July 6 19:28 UTC) no QPOs were detected during the remainder of the count rate and soft color plateau. However, as the soft color returned to its pre-plateau values, a weak QPO appeared around 9 Hz, which we suspect was a type-C QPO. We did not see any indications for high-frequency QPOs in our power density spectra.

\subsection{Radio flare}\label{sec:radio}

\begin{figure}[t]
\centerline{\includegraphics[width=8.5cm]{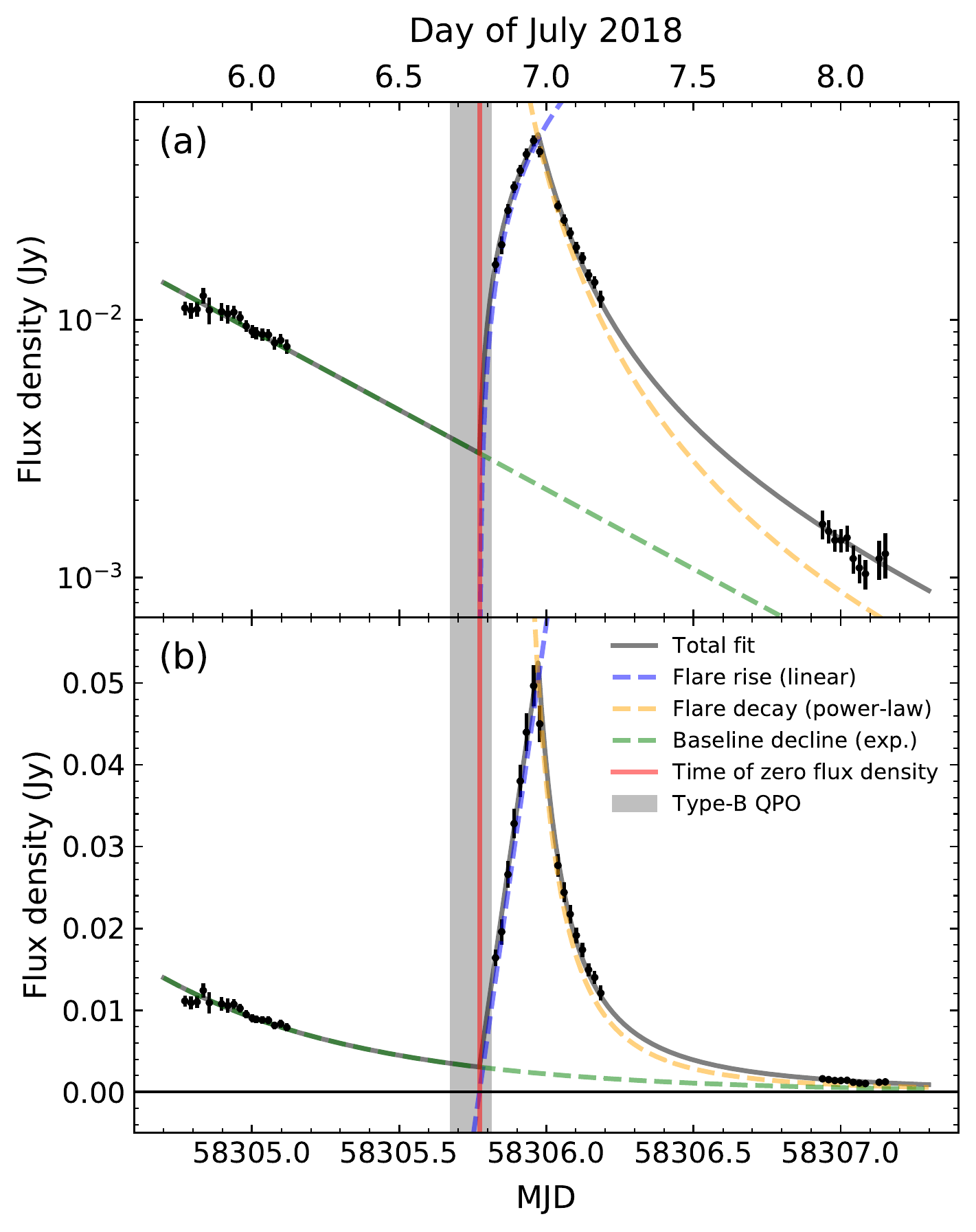}}
\caption{AMI-LA radio light curve (15.5 GHz) of \maxi\ (black dots). The time resolution is 30 minutes. Both panels show the same data, but with different y-axis scales.  The gray shaded areas mark the time period during which the type-B QPO was present. The radio data were fit with a model consisting of three functions: an exponential function for the baseline decay (green dashed line), a linear function for the rise (blue dashed line), and a power law for the flare decay (orange dashed line). The linear component was only applied between the start time of the flare and break time between the linear rise and power-law decay, while the power law was only applied after the break time (MJD 58,305.973$\pm$0.003 or July 6 23:21 UTC). The baseline component contributed at all times. The dark gray line shows the total contribution from all three components combined. We note that in this model the flare decay still dominates the flux density in the third data segment. The start time of the radio flare is defined as an extrapolation of the rise back to zero flux density, using the linear fit to the rise (blue dashed line). This start time is marked with a vertical red line (MJD 58,305.773$\pm$0.006 or July 6 18:33 UTC).}
\label{fig:flare}
\end{figure}

The AMI-LA observations covered part of the rapid state transition described in Section \ref{sec:xray}. The radio data, which are shown in detail in Figure \ref{fig:flare} and also in Figure \ref{fig:transition}c, were obtained between July 5 $\sim$18:31 UTC and July 8 at $\sim$03:39 UTC (MJD 58,304.77201--58,307.15169) in three separate segments.  \citet{brmofe2018} reported a sharp drop in the radio flux density ($S_\nu$) between July 2 and July 5, from  $\sim$40 mJy to $\sim$10 mJy.  As can be seen from Figure \ref{fig:flare} the flux density was still declining during the first segment on July 5/6. However, during the second segment (July 6/7) a strong radio flare occurred, reaching a maximum flux density of $\sim$50 mJy on July 6 at 22:58 UTC (MJD 58,305.95697). \citet{brfemo2020} showed that this radio flare  was the result of the launch of long-lived superluminal ejecta.  The start of the radio flare took place during the gap between the first and second data segments.
In the  third segment (July 7/8) the evolution appeared to be a continuation of the trend seen during the first segment, with the flux density declining on a similar time scale. This suggests that the occurrence of the radio flare did not significantly affect the overall decline in radio flux density.

Usually, the start of a radio flare is defined as the moment at which the flux starts to deviate significantly from the persistent flux \citep[see, e.g.,][]{gacofe2004}. This can be approximated by finding the intersection of the baseline decline (seen in the first and third data segments) and the rise of the radio flare. However, this definition of the start time of the flare is quite sensitive to the baseline flux level, which is not necessarily related to the flux from the radio flare. Therefore, we define an alternative start time by extrapolating the rise of the radio flare back to $S_\nu=0$ mJy. 
We fit the entire radio data set with three components, which are shown in Figure \ref{fig:flare} (see the caption for details): a linear function for the rise of the flare, a power law for the flare decay, and an exponential function for the baseline decay.  The combined fit (gray line) results in a reduced $\chi^2$ of 0.86 for 37 degrees of freedom. Extrapolating the linear fit (blue dashed lines in Fig.\ \ref{fig:flare}) back to  $S_\nu=0$ mJy, we find a start time of MJD 58,305.773$\pm$0.006 (or July 6 18:33 UTC, with a statistical uncertainty of $\sim$10 minutes). This time is marked as a vertical red line in Figure \ref{fig:flare} and in all panels of Figure \ref{fig:transition}. The estimated flare start is $\sim$2--2.5 hr later than the time at which the type-B QPO appeared. However, it still falls within the shaded gray area in Figures \ref{fig:transition} and \ref{fig:flare} that represents the presence of the type-B QPO.

\section{Discussion}\label{sec:discussion}

We have presented {\it NICER} X-ray and AMI-LA radio coverage of a rapid hard-to-soft state transition in the black hole X-ray transient \maxi. During this state transition a switch from type-C to type-B QPOs was seen in the power spectra, together with a drop in the strength of the broadband X-ray variability (Fig.\ \ref{fig:qpos}) and the start of a brief flare in the 7--12 keV band  (Fig.\ \ref{fig:transition}f). Soon after the rapid state transition a strong radio flare was observed, which was shown to be the result of the launch of relativistic ejecta by \citet{brfemo2020}. Assuming that the X-ray flare and the dramatic change in the X-ray timing properties mark the time of the ejection event---perhaps similar to how strong X-ray spikes often mark ejection events in the BH LMXB GRS 1915+105 \citep{eimamo1998,midhch1998}---the start of the radio flare is delayed with the respect to the launch of the ejecta by $\sim$2--2.5 hr. Since the AMI in-band spectral index indicates that the flare was optically thin throughout (J.\ Bright et al.\ in preparation), the most likely explanation for the observed delay is that it represents the time it takes for the ejecta to catch up with slower moving jet material that was ejected previously, or the time it takes for the ejecta to run into the interstellar medium. Both would result in optically thin radio emission due to internal shocks \citep{kasusp2000,mirutu2009,jafeka2010,ma2014}.

\citet{fehobe2009}, \citet{misial2012}, and \citet{rutemi2019} looked in detail at possible connections between the occurrence of radio flares and changes in the X-ray timing properties in several transient  BH LMXBs: XTE J1550--564, XTE J1859+226, GX 339--4, H1743--322,  XTE J1752--223, and MAXI J1535--571.  Despite some cases in which a radio flare occurred close in time (within $\sim$half a day to days) to the disappearance of type-C QPOs and/or the appearance of type-B QPOs, the authors concluded that there was not enough evidence to directly tie discrete jet ejections to a change in X-ray timing properties.  As the authors already noted, however, their data sets suffered from gaps (up to several days long) in the radio and/or X-ray coverage, limiting their ability to either confirm or rule out a connection between the phenomena.

The {\it NICER} and AMI-LA data sets for \maxi\ that we used in this paper present a significant improvement over the data sets analyzed by \citet{fehobe2009}, \citet{misial2012}, and \citet{rutemi2019}. Most importantly, in X-rays we were able to follow the rapid evolution of the type-C QPOs until they disappeared and we captured the moment at which a narrow type-B QPO appeared around $\sim$4.5 Hz; in the previously studied sources this switch between type-C and type-B QPOs always happened during gaps in the X-ray coverage.  Moreover, despite missing the start of the radio flare, the otherwise good radio coverage allowed us to narrow down its start time to a period of  $\sim$2--2.5 hr after the appearance of the type-B QPO. What sets the \maxi\ data set apart from the data sets discussed above is the relative accuracy with which we were able to determine both the start of the radio flare and the appearance of the type-B QPO.
Although we cannot rule out that the radio flare and the switch in QPO type are not physically related, the short ($\sim$2--2.5 hr) time delay is the strongest empirical evidence to date to suggest that they are, in fact, connected. This does not, of course, necessarily imply that radio flares are only produced when type-B QPOs appear.

It is important to point out that the data sets studied by \citet{fehobe2009}, \citet{misial2012}, and \citet{rutemi2019},  are not inconsistent with there being a strong connection between radio flares and the appearance of type-B QPOs. However, for their behavior to be consistent with this we have to assume that type-B QPOs were present during gaps in the X-ray data prior to the detection of the radio flares. This is not an unreasonable assumption: while the phase during which type-B QPOs are observed in a given source can be a few days long, type-B QPOs themselves are often short-lived \citep[minutes to tens of minutes; see, e.g.,][]{cabeho2004} and can easily be missed.  

Finally, the appearance of the type-B QPO coincided with the start of a small flare in the 7--12 keV band (see Fig.\ \ref{fig:transition}f)---a detailed spectral analysis of the rapid spectral evolution during this phase will be presented in future work. This flare had a duration between 3.2 and 4.6 hr (the presence of a data gap prevents a more accurate estimate). It has been proposed that the base of a compact jet in the hard state can produce X-rays through synchrotron emission and inverse Compton scattering \citep{manowi2005}. Although  \citet{manowi2005} do not specifically discuss X-ray emission from discrete jet ejections, the presence of the short 7--12 keV flare preceding the radio flare suggests that these ejections may produce (hard) X-rays as well, either at the time of their launch or very soon after. This would also strengthen the case for a jet-based origin of type-B QPOs, as suggested by \citet{mocahe2015} based on their dependence on viewing angle.

\section{Conclusions}

The dense {\it NICER} coverage of the rapid state transition in MAXI J1820+070 has allowed us to look for transient phenomena in X-rays that appear to be associated with discrete jet ejections. We have identified two such phenomena: the appearance of a type-B QPO and the occurrence of a flare in the 7--12 keV band, both of which lasted for approximately four hours and started within $\sim$2--2.5 hr of the inferred onset of a radio flare that was observed with AMI-LA. They may represent the most promising observational X-ray markers of discrete jet ejections in transient BH LMXBs.

\acknowledgments This work was supported by NASA through the \textit{NICER} mission and the Astrophysics Explorers Program. This research has made use of data, software and/or web tools obtained from the High Energy Astrophysics Science Archive Research Center (HEASARC), a service of the Astrophysics Science Division at NASA/GSFC and of the Smithsonian Astrophysical Observatory's High Energy Astrophysics Division. We thank the Mullard Radio Astronomy Observatory staff for scheduling and carrying out the AMI-LA observations. The AMI telescope is supported by the UK Science and Technology Facilities Council, and the University of Cambridge. J.H. acknowledges financial support from NASA grant 80NSSC19K1466. D.A.\ acknowledges support from the Royal Society. A.L.S. is supported by an NSF Astronomy and Astrophysics Postdoctoral Fellowship under award AST-1801792. T.M.B.\ acknowledges financial contribution from the agreement ASI-INAF n.2017-14-H.0.

\facility{\textit{NICER}, AMI-LA}

\end{document}